\documentclass[preprint,aps,showpacs]{revtex4}
\usepackage{dcolumn}
\usepackage{bm}
\usepackage{epsfig}
\usepackage{epsf}

\begin{document}
\title{Abnormal phenomena in a one-dimensional periodic structure containing left-handed materials}
\vspace{2in}
\author{Liang Wu$^{1}$, Sailing He$^{1,2}$ and Linfang Shen$^{1}$}

\address{$^{1}$
Center for Optical and Electromagnetic Research,\\
State Key Laboratory for Modern Optical Instrumentation, \\
Zhejiang University, \\
Yu-Quan, Hangzhou 310027, P. R. China\\
$^{2}$Division of Electromagnetic Theory, Alfven Laboratory,  Royal Institute of Technology, \\
S-100 44 Stockholm, Sweden }
\date{\today}

\begin{abstract}

The explicit dispersion equation for a one-dimensional  periodic
structure with alternative layers of left-handed material (LHM)
and right-handed material (RHM) is given and analyzed. Some
abnormal phenomena such as spurious modes with complex
frequencies, discrete modes and photon tunnelling modes are
observed in the band structure. The existence of spurious modes
with complex frequencies is a common problem in the calculation of
the band structure for such a photonic crystal. Physical
explanation and significance are given for the discrete modes
(with real values of wave number) and photon tunnelling
propagation modes (with imaginary wave numbers in a limited
region).

\end{abstract}

\pacs{78.20.Ci, 42.70.Qs, 78.20.-e, 03.65.Ge,}




\maketitle

Left-handed materials (LHM)  with negative permittivity and
negative permeability, which were first suggested in
\cite{veselago}, have attracted much attention recently
\cite{smith,shelby,pendry1} thanks to the experimental realization
of such materials \cite{science} and the debate on the use of a
LHM slab as a perfect lens to focus both propagating waves and
evanescent waves \cite{pendry,prl,surface}.

Multi-layered structures containing LHM have been investigated
through calculating the transmittance or reflectance of the
structure \cite{matrix,bragg}. Effects of photon tunnelling and
Bragg diffraction have been observed in these works. In the
present paper, we consider  a  one-dimensional (1D) photonic
crystal formed by alternative layers of LHM and RHM.   The band
structure is investigated by analyzing the explicit dispersion
equation.
 Some abnormal phenomena such as spurious modes with complex frequencies, discrete modes and photon tunnelling modes are found and explained in the present paper.

\begin{figure}
\includegraphics[width=3.5in]{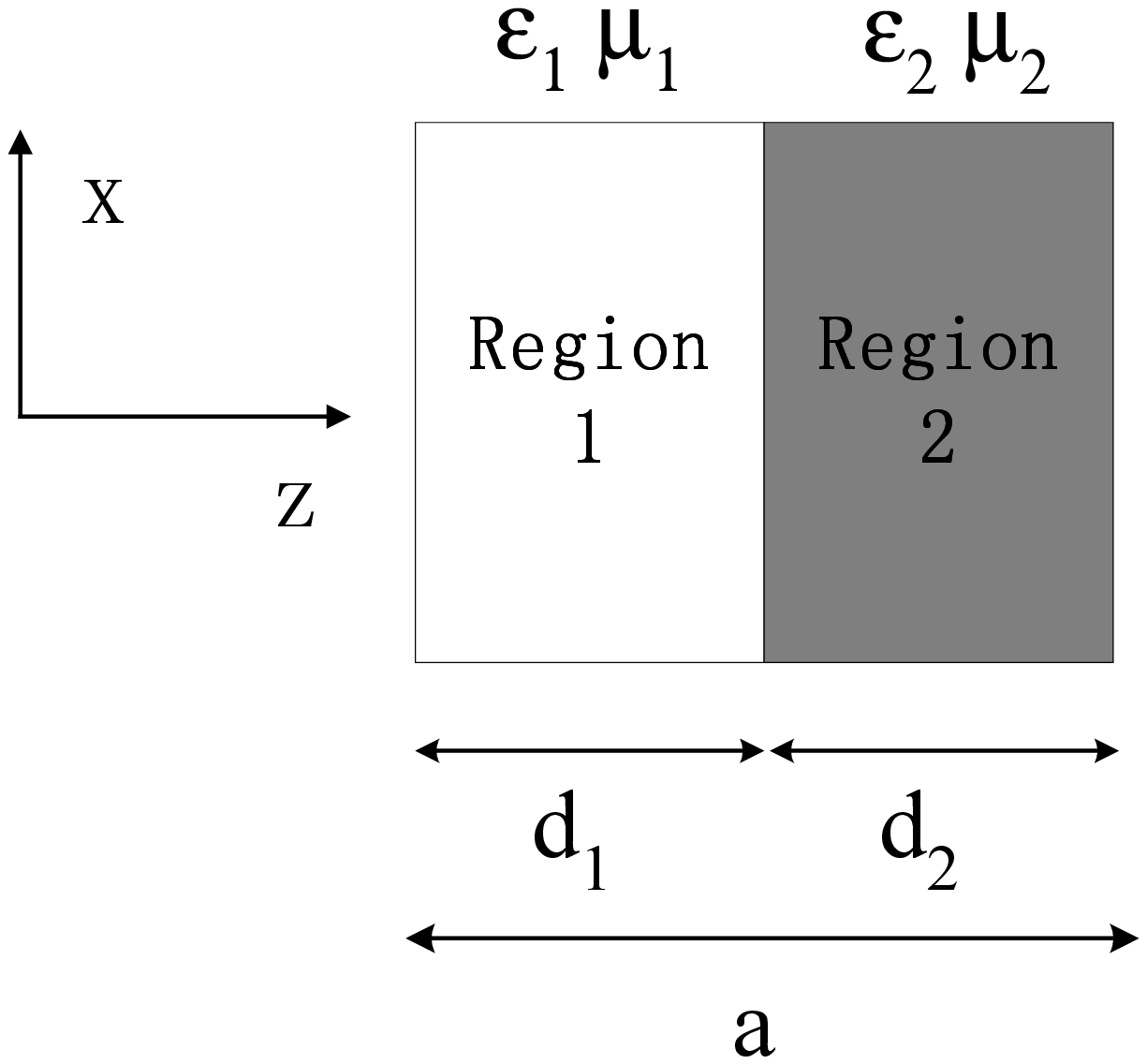}
\caption{\label{fig:fig1} A 1D periodic structure consisting of
alternate layers of RHM and LHM inclusions.}
\end{figure}

Consider a 1D periodic structure with alternating layers
$(\epsilon_1,\mu_1)$ and $(\epsilon_2 ,\mu_2)$ as shown in
Fig.~\ref{fig:fig1}.  $d_1$ and $d_2$ are the widths of the two
inclusion layers respectively, and $a=d_1+d_2$ is the period. Note
that the conservation of energy requires that $\epsilon_i\mu_i>0$,
$i=1,2$. The corresponding refractive index is given by $n_i =
\pm \sqrt{\epsilon_i\mu_i }$ (negative sign for LHM). We consider
an off-line propagation of monochromatic electromagnetic field
(with time-dependence $e^{-i\omega t}$) in a periodic structure
with off-line wave vector $\beta$ along the $x$ axis.  For the
E-polarization case, one has
\begin{eqnarray}
\left\{
\begin{array}{lll}
E_{1y}=e^{i\beta x}(e^{ik_1 z}+Ae^{-ik_1 z}) \\
H_{1x}=\frac{-k_1}{\omega\mu_1}e^{i\beta x}(e^{ik_1 z}-Ae^{-ik_1 z}) \\
H_{1z}=\frac{\beta}{\omega\mu_1}e^{i\beta x}(e^{ik_1 z}+Ae^{-ik_1
z})
\end{array}\right. \quad \mbox{in region 1},
\end{eqnarray}
and
\begin{eqnarray}
\left\{
\begin{array}{lll}
E_{2y}=e^{i\beta x}(Be^{ik_2 z}+Ce^{-ik_2 z})   \\
H_{2x}=\frac{-k_2}{\omega\mu_2}e^{i\beta x}(Be^{ik_2 z}-Ce^{-ik_2 z})  \\
H_{2z}=\frac{\beta}{\omega\mu_2}e^{i\beta x}(Be^{ik_2 z}+Ce^{-ik_1
z})
\end{array}\right. \quad \mbox{in region 2},
\end{eqnarray}
where $k_i$ is the component of the wave vector along the $z-$axis
in region $i$ ($i=1,2$), i.e., $k_i^2=
\omega^2\epsilon_i\mu_i/c^2-\beta^2$. Here $c$ is the speed of
light in vacuum.

The tangential electric and magnetic fields should be continuous
at $z=0$, i.e.,
\begin{eqnarray}
\left\{
\begin{array}{ll}
E_{1y}(z=0^-)=E_{2y}(z=0^+) \\
H_{1x}(z=0^-)=H_{2x}(z=0^+),
\end{array} \right.
\end{eqnarray}
To obtain the dispersion relation for this 1D photonic crystal, we
need to use the following periodic conditions according to the
Bloch theorem,
\begin{eqnarray}
\left\{
\begin{array}{ll}
E_{2y}(z=d_2)=E_{1y}(z=-d_1)e^{iqa}  \\
H_{2x}(z=d_2)=H_{1x}(z=-d_1)e^{iqa},
\end{array} \right.
\end{eqnarray}
where $q$ is in the first Brillouin zone $-\frac{\pi}{a} \le q\le
\frac{\pi}{a}$ . Substituting systems (1) and (2) into systems (3)
and (4), we obtain the following dispersion relation for the
E-polarization case,
\begin{eqnarray}
\cos(k_1d_1)\cos(\tilde{n}k_1d_2)-\frac{{\tilde{n}}^2+\mu^2}{2\tilde{n}\mu}\sin(k_1d_1)\sin(\tilde{n}k_1d_2)=\cos(qa)
\label{eq:Ep},
\end{eqnarray}
where $\tilde{n}=k_2/k_1=\sqrt{n^2(k_1^2+\beta^2)-\beta^2}/k_1$,
$\mu=\mu_2/\mu_1$, and $n=n_2/n_1$ is the index ratio of the two
media.

Similarly, we can derive the following dispersion relation for the
H-polarization case,
\begin{eqnarray}
\cos(k_1d_1)\cos(\tilde{n}k_1d_2)-\frac{{\tilde{n}}^2+\epsilon^2}{2\tilde{n}\epsilon}\sin(k_1d_1)\sin(\tilde{n}k_1d_2)=\cos(qa)
\label{eq:Hp},
\end{eqnarray}
where $\epsilon=\epsilon_2/\epsilon_1$.

From analytical equations (5) and (6), we can find out how $k_1$
depends on $q$. The corresponding dispersion relation between
$\omega$ and $q$ can then be obtained according to $\omega^2=c^2
(k_1^2+\beta^2) /(\epsilon_1\mu_1)$ for each fixed $\beta $
($\beta=0$  is associated to the in-line propagation, i.e, the
case when the propagation direction is along the normal of the
material interfaces).


When the index ratio $n<0$, the solution to the dispersion
equation and the associated band structure have quite different
behaviors  as compared to those for the usual case when $n>0$. For
simplicity, we first illustrate this for the case when $\beta=0$.
In this case,
$\frac{{\tilde{n}}^2+\mu^2}{2\tilde{n}\mu}=\frac{{\tilde{n}}^2+\epsilon^2}{2\tilde{n}\epsilon}$and
the dispersion equations for the E-polarization and the
H-polarization have the same form (as expected).

\begin{figure}
\includegraphics[width=3.5in]{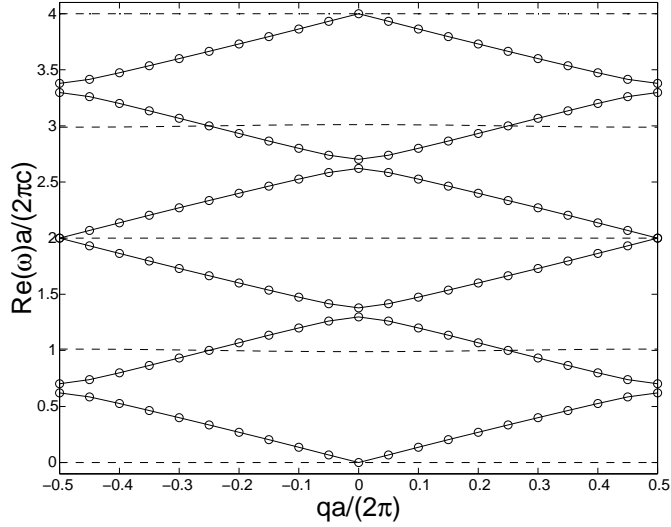}
\caption{\label{fig:fig2} The band structure for the in-line
propagation (i.e., $\beta=0$) with $\epsilon_1=1, \mu_1=1, n=-2.5,
\mu_2=-2, d_1=d_2=0.5a$. The circles show the real $\omega $,
while the dashed lines indicate the real parts of the complex
$\omega $. }
\end{figure}

{\bf Spurious modes with complex $\omega $.}   When one of the
constitutive  media is of LHM (i.e., $n<0$),  the analytical
dispersion equation may have some complex solutions for $k_1$.
These complex values of $k_1$ lead to complex $\omega$ since
$\omega^2=c^2(k_1^2+\beta^2)/(\epsilon_1\mu_1) $.
Fig.~\ref{fig:fig2} gives the in-line (i.e., $\beta=0$) band
structure when $n=-2.5, \mu=-2, d_1=d_2=0.5a$. The real part of
$\omega$ varies little as $q$ varies from $-\pi/a$ to $\pi/a$. The
problem of complex $\omega $ will also appear when $\beta \ne 0$
(i.e., the case of off-line propagation). Obviously, these complex
$\omega $ have no physical significance since physically the
frequency $\omega $ must have a real value.

One would meet a similar problem of complex $\omega $ in the
calculation of the band structure using other methods such as the
plane wave expansion method (one needs to remove these spurious
complex $\omega $ in the band structure). In the plane wave
expansion method, the $\omega $ values  are determined from the
following eigenvalue equation, $\frac{1}{\mu}{\bf
\nabla}\times(\frac{1}{\epsilon}{\bf\nabla}\times{\bf
H})=\frac{\omega^2}{c^2}{\bf H}$. The existence of complex $\omega
$ indicates that the operator $\frac{1}{\mu}{\bf
\nabla}\times(\frac{1}{\epsilon}{\bf\nabla})$ is not a Hermit
operator or has no equivalent Hermit operator. One can also expect
that these complex $\omega$ (associated with an exponential
increasing factor of time) will make the finite-difference time
domain (FDTD) method (with a periodic boundary condition)
divergent in the calculation of the band structure of a photonic
crystal consisting of alternate RHM and LHM inclusions (with
constant permittivity and permeability) \cite{ PRE, chan,wu}. We
have verified numerically with the plane wave expansion method and
the FDTD method the existence of the spurious modes with complex
$\omega $.

The fields associated with the spurious complex $\omega$ satisfy
Maxwell's equations, the boundary condition and the enforced
periodic condition.
 Therefore, an additional requirement that the frequency $\omega$ must
be real should be applied to make the solution significant in
physics.

In the rest of the paper, we consider only situations when $\omega
$ is real.

When $\beta = 0$, a complex solution of $k_1$ always leads to a
spurious complex $\omega $. However,  when $\beta \ne  0$ some
complex solutions of $k_1$  may correspond to real $\omega $ and
thus  have physical significance. Since $\omega^2=c^2
(k_i^2+\beta^2) /(\epsilon_i\mu_i)$, a complex solution of $k_i
\equiv k_{iR}+jk_{iI}$ (with $k_{iI}\ne 0$) should satisfies the
following conditions in order to make $\omega$ real,
\begin{eqnarray}
k_{iR}=0, \quad k_{iI}^2<\beta^2. \label{eq:con}
\end{eqnarray}
In other words, an imaginary solution of $k_1$ can have  physical
significance when $\beta \ne 0$ (i.e., the off-line propagation
case).  Imaginary solutions of $k_1$ correspond to evanescent
waves. In a periodic structure consisting of alternate RHM and LHM
layers, the special property that the evanescent waves (in both
directions) decrease in the RHM layer and increase in the LHM
layer \cite{pendry} or vice versa permits the evanescent waves to
propagate in the periodic structure (i.e.,  a propagation mode,
for which the energy density does not decay over a period; we will
call it a photon tunnelling mode and will discuss it in more
details below). In the periodic structure composed solely by RHM
or LHM, complex solutions of $k_1$ will not occur since evanescent
waves can not form a propagation mode in these structures.


As discussed before,  imaginary solutions of $k_1$ may have
physical significance when $\beta\ne 0$. This will be illustrated
by the abnormal photon tunnelling modes in the band structure for
a photonic crystal consisting of alternate RHM and LHM layers.
Another abnormal (and interesting) phenomenon is that some
discrete modes (with real values of wave number) are also found in
the band structure.

To simplify the analysis, we consider a special case when $n=-1$
for the E-polarization. Then $\tilde{n}=-1$ (independent of
$\beta$) and Eq.~(\ref{eq:Ep}) is simplified to
\begin{eqnarray}
\cos(k_1d_1)\cos(k_1d_2)-\frac{1+\mu^2}{2\mu}\sin(k_1d_1)\sin(k_1d_2)=\cos(qa).
\label{eq:we}
\end{eqnarray}
One notices that the above equation for $k_1$ is independent of
$\beta$.We discuss situations when $\mu\ne -1$ (so that the RHM
and LHM layers have different values of impedance). It is
particularly interesting when the two inclusion layers have the
same  width (i.e., $d_1=d_2=0.5a$) since in this case only
discrete modes and photon tunnelling modes exist (conventional
propagating band disappears in the band structure).

\begin{figure}
\includegraphics[width=3.5in]{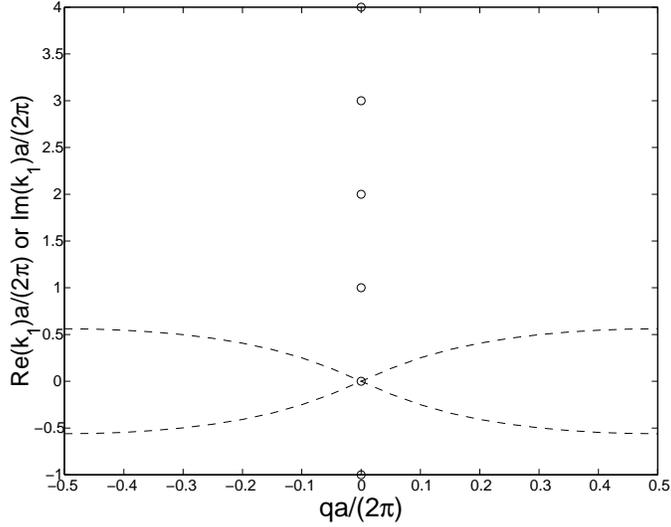}
\caption{\label{fig:fig3} The solutions (with physical
significance) of $k_1$ to Eq. (8) with $\epsilon_1=1, \mu_1=1,
n=-1, \mu_2=-2, d_1=d_2=0.5a$. The circles indicate the real
solutions, which exist discretely at $q=0$. The dashed lines show
the imaginary solutions. The imaginary solutions are continuously
distributed in a region from $-0.561j\times 2\pi/a$ to
$0.561j\times 2\pi/a$. }
\end{figure}

{\bf Discrete modes.} Since
$\cos(k_1d_1)\cos(k_1d_2)+\frac{1+\mu^2}{2|{\mu}|}\sin(k_1d_1)\sin(k_1d_2)\ge
1$ (when $k_1$ is real),   real solutions of $k_1$ to Eq. (8)
exist only when $q=0$ (notice that $-\pi/a\le q\le\pi/a$), which
corresponds to discrete solutions $k_1=2N\pi/a$, $N=0, \pm 1, \pm
2, \cdots$. Since these discrete solutions satisfy the transparent
condition of $k_1d_1=N\pi,\, k_2d_2=-N\pi$ (one can easily show
that  each layer is transparent to neighboring layers at these
discrete frequencies for a fixed $\beta $), the field with this
wave vector will surely pass through the periodic structure and
becomes a propagating mode. On the other hand,
 fields with any other wave vector (i.e., any other frequency for
 the fixed $\beta $) satisfies the Bragg reflective condition
 $k_1d_1+k_2d_2=0$ and can not propagate in this periodic structure. Therefore, only
those discrete propagation modes exist. As an example, the
discrete real solutions of $k_1$ to Eq. (8) are shown by the
circles in Fig.~\ref{fig:fig3} (with $\mu=-2$).  This abnormal
phenomenon (existing for any $\beta $) can be utilized to make a
very narrow filter (with no sidelope, which is different from any
conventional type of filters).

{\bf Photon tunnelling modes.} From Eq. (8) one sees that only
complex $k_1$ can be the solutions if $q\ne 0$. Then according to
condition (7) the solutions $k_1$ with physical significance must
be imaginary.
  The expression on the left-hand
side of Eq.~(\ref{eq:we}) always decreases when $k_{1I}$ increases
or decreases away from $0$ (a symmetric function of $k_{1I}$ with
a maximum at $k_{1I}=0$). Therefore, only two conjugate imaginary
solutions exist. The dashed lines in  Fig.~\ref{fig:fig3} show the
imaginary solutions  of $k_1$ to Eq. (8) for an example. When $q
(\ne 0)$ varies in the first Brillouin zone,  the two conjugate
imaginary solutions vary continuously in a region from
$-0.561j\times 2\pi/a$ to $0.561j\times 2\pi/a$. Note that this
figure is independent of $\beta $ for this special case. For
off-line waves with $\beta^2>k_{iI}^2$, the imaginary solutions
correspond to special propagation modes$-$photon tunneling modes
(as discussed before). For this example, evanescent modes with $
k_{1I}$ outside the region $[-0.561j\times 2\pi/a, 0.561j\times
2\pi/a]$ can not propagate inside the periodic structure. From
Fig.~\ref{fig:fig3} one can also see that $k_{1I}$ approaches 0
when $q$ approaches 0. Thus, any $\beta\ne 0$ can permit the
existence of some photon tunnelling modes since $k_{1I}$ can be
arbitrarily small to satisfy $k_{1I}^2 < \beta^2$.  From the above
discussion, one sees that for any $\beta\ne 0$ there always exist
some evanescent modes which can propagate inside the periodic
structure while some other evanescent modes can not (one notices
that in the case of a perfect lens \cite{pendry} all the
evanescent waves can go through the lens). This limitation of the
photon tunnelling modes is due to the enforced periodic condition.
Thus, the problem of square non-integrable field mentioned in
\cite{prl} will not occur in the present situation (this is also
true for the next example when $d_1\ne d_2$).

Both discrete modes and photon tunnelling modes exist when $\beta
\ne 0$. However,  when $\beta=0$ (as in the above example), which
corresponds to normally incident waves (i.e., the in-line
propagation case), only discrete modes  can exist (this means that
only fields at discrete frequencies can exist in the structure).

\begin{figure}
\includegraphics[width=3.5in]{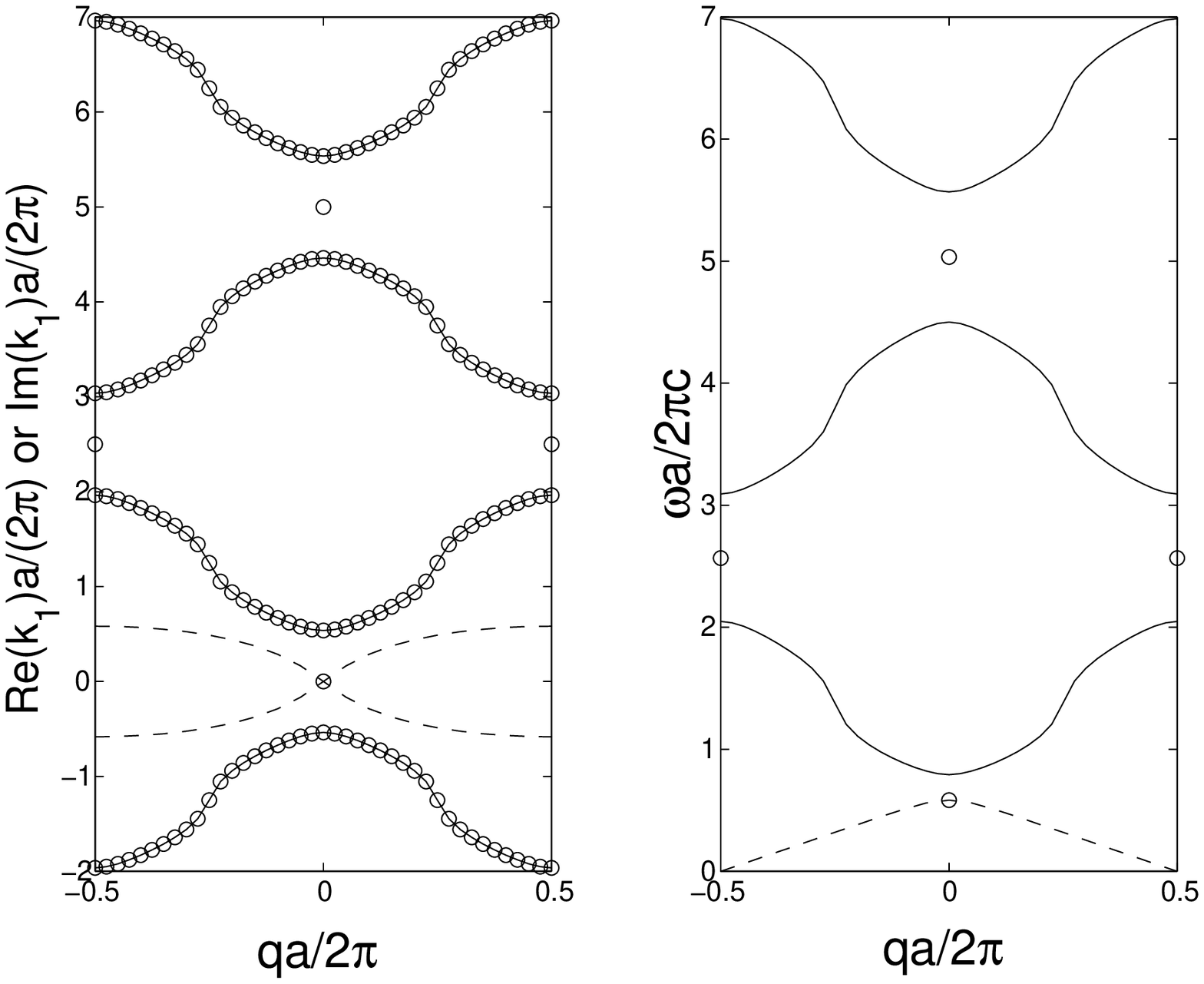}
\caption{\label{fig:fig4}  Off-line case with $d_1 \ne d_2$.
The two inclusion layers have $\epsilon_1=1, \mu_1=1, n=-1, \mu_2=-2, d_1=0.4a$, and $d_2=0.6a$.
(a) The solutions of $k_1$ with physical significance.
The circles show the real solutions. The real solutions have
discrete values (at  $q=0$ and $q= \pm \pi/a $) as well as
continuous bands. The dashed lines indicate the imaginary parts of
the imaginary solutions. (b) The band structure when
$\beta=0.583\times 2\pi/a$. The dashed lines show the frequencies
of the photon tunnelling modes. The circles indicate  the discrete
frequencies of the discrete modes. The solid lines show the
continuous bands. }
\end{figure}

In the above discussion  the case when $d_1=d_2$ is considered.
Fig.~\ref{fig:fig4}(a) shows the solutions of $k_1$ for an example
when $d_1\ne d_2$. From this figure one sees that non-discrete
real $k_1$ solutions (corresponding to some continuous bands shown
in Fig.~\ref{fig:fig4}(b) exist besides some discrete modes and
photon tunnelling modes. The discrete solutions are at
$k_1=(2.5+5N) \times 2\pi/a $ for $q=\pm \pi/a$ and at
$k_1=5N\times 2\pi/a $ for $q=0$, $N=0, \pm 1, \pm 2, $$бн$. They
are located in the forbidden band-gaps (see Fig.~\ref{fig:fig4}(b)
for the corresponding band structure for a given $\beta$).


In the above we have discussed only the ideal situations when the
periodic structure consists of ideal materials with lossless and
non-dispersive negative $\epsilon$ and negative  $\mu$ ($\epsilon$
and $\mu$ are the relative permittivity and permeability of the
two inclusion media). In a non-ideal situation with material loss
or dispersion the abnormal phenomena such as spurious modes with
complex $\omega $, discrete modes and photon tunnelling modes can
also be expected to occur for the following reasons.

If the inclusion media are dispersive, $\epsilon$ and $\mu$ will
be frequency-dependent. Since the solutions can be obtained from
Eq. (5) in a point-by-point manner, they are correct locally
(i.e., in a small frequency range where the frequency-dependence
of $\epsilon$ and $\mu$ can be neglected). Thus, the abnormal
phenomena can still exist in certain small frequency region.
Furthermore, the frequency in the band structure is scalable and
thus we can scale it to a frequency at which the material
parameters have desired values in order to observe these abnormal
phenomena.

If the LHM inclusion is lossy the relative $\epsilon$ or $\mu$ has
a very small positive imaginary part (e.g., $10^{-5}j$; as claimed
in some papers that LHM must have some loss \cite{prl}). The
abnormal phenomena can still occur since the dispersion equation
(Eq. (5)) will not give any rapid change when a very small
imaginary part is added to $\epsilon$ or $\mu$. Note that for the
photon tunnelling modes $q$ should be complex (corresponding to a
pseudo-periodic field with a damping factor along the propagation
direction) when $\epsilon$ or $\mu$ is complex. Since the
imaginary parts of these parameters are very small, Eq. (5) is
still valid when all the parameters (including $\epsilon$, $\mu$
and $q$) are replaced with their real parts.

In conclusion,  we have found some abnormal phenomena such as
spurious modes with complex $\omega $, discrete modes and photon
tunnelling modes by analyzing the  explicit dispersion equation
for a 1D periodic structure with alternating LHM and RHM layers.
One will meet the problem of complex $\omega $ in the calculation
of the band structure using other methods such as the plane wave
expansion method and the FDTD method. The physical significance of
the discrete modes  and photon tunnelling modes has been
explained. The discrete modes can be utilized to make a very
narrow filter with no sidelope. For an off-line wave there exist
some photon tunnelling propagation modes while some other
evanescent modes can not propagate inside the periodic structure.

{\bf Acknowledgement.} The partial support of National Natural
Science Foundation of China (under a key project grant; grant
number 90101024)  is gratefully acknowledged.


\begin{references}
\bibitem{veselago} V.G. Veselago, Sov. Phys. Usp. {\bf 10}, 509
(1968).
\bibitem{smith} D.R. Smith, W.J. Padilla, D.C. Vier, S.C.
Nemat-Nasser, and S.Schultz, Phys. Rev. Lett. {\bf 84}, 4184
(2000).
\bibitem{shelby} R.A. Shelby, D.R. Smith, S.C. Nemat-Nasser, and
S. Schultz, Appl. Phys. Lett. {\bf 78}, 489 (2001).
\bibitem{pendry1} J.B. Pendry, A.J. Holden, W.J. Stewart, and I.
Youngs, Phys. Rev. Lett. {\bf 76}, 4773 (1996).
\bibitem{science}
R.A. Shelby,  D.R. Smith and S. Schultz, {\it Science} {\bf 292},
77 (2001).
\bibitem{prl} N. Garcia and M. Nieto-Vesperinas, Phys. Rev. Lett., {\bf 88}, 207403 (2002).
\bibitem{surface} Michael W. Feise, Peter J. Bevelacqua, and John B. Schneider, Phys. Rev. B, {\bf 66}, 035113 (2002).
\bibitem{pendry} J. B. Pendry, Phys. Rev. Lett. {\bf 85}, 3966
(2000).
\bibitem{matrix} Z. M. Zhang and C. J. Fu, Appl. Phys. Lett. {\bf
80}, 1097 (2001).
\bibitem{bragg} J. Gerardin and A. Lakhtakia, Mirc. and Opti.
Tech. Lett. {\bf 34}, 409 (2002).

\bibitem{PRE} R. W. Ziolkowski and E. Heyman, Phys. Rev. E, {\bf
64}, 056625 (2001).

\bibitem{chan} C. T. Chan, Q. L. Yu, and K. M. Ho, Phys. Rev. B, {\bf 51}, 16635 (1995).
\bibitem{wu} L. Wu and S. He, J. Appl. Phys., {\bf 91}, 6499 (2002).
\end{references}
\end{document}